\documentclass[numrefs]{wiley-article}
\usepackage{graphicx}
\usepackage[space]{grffile}
\usepackage{latexsym}
\usepackage{textcomp}
\usepackage{longtable}
\usepackage{tabulary}
\usepackage{booktabs, array, multirow}
\usepackage{amsfonts,amsmath,amssymb}
\usepackage{url}
\usepackage{natbib}
\usepackage{hyperref}
\usepackage[labelsep=period]{caption}
\hypersetup{colorlinks=false,pdfborder={0 0 0}}
\usepackage{etoolbox}
\newif\iflatexml\latexmlfalse

\AtBeginDocument{\DeclareGraphicsExtensions{.pdf,.PDF,.eps,.EPS,.png,.PNG,.tif,.TIF,.jpg,.JPG,.jpeg,.JPEG}}

\usepackage[utf8]{inputenc}
\usepackage[english]{babel}

\usepackage{siunitx}
\usepackage{comment}

\iflatexml
\else
\paperfield{Field of the paper}
\abbrevs{ED, Emergency Department; IP, Inpatient; LoS, Length of Stay; HCCIS, Health Care Cost Information System}
\corraddress{Nathan Adeyemi, Mechanical and Industrial Engineering Department, Northeastern University, Boston, MA, USA}
\corremail{adeyemi.n@northeastern.edu}

\papertype{Original Article}

\title{Policy Interventions to Improve Inpatient Mental Healthcare Access: A Discrete Event Simulation Study}

\author[1]{Nathan O. Adeyemi}
\author[1]{Kayse Lee Maass}
\author[2]{Amanda M. Graham}
\author[3]{Kalyan S. Pasupathy}

\affil[1]{Mechanical and Industrial Engineering Department, Northeastern University, Boston, Massachusetts, USA}
\affil[2]{Mayo Clinic, Rochester, Minnesota, USA}
\affil[3]{Department of Biomedical and Health Information Sciences, University of Illinois at Chicago, Chicago, Illinois, USA}

\runningauthor{Adeyemi et al.}

\begin{document}

\maketitle
\selectlanguage{english}
\begin{abstract}
For a large portion of mental health patients, the Emergency Department is the first point of contact when in crisis and in need of urgent acute care. Unfortunately, those who have already received an admission disposition may wait hours or days after before being placed in a psychiatric inpatient (IP) care unit. Known as ED boarding, one primary contributor is the inability to locate an available IP bed for transferred patients. In this study, we develop a discrete event simulation modeling patients arriving at numerous EDs throughout the region, as they're either internally placed in a psychiatric IP care unit or externally transferred to an IP unit located outside the ED they originally arrived. This simulation is then used to investigate the effect of three proposed interventions to the IP bed placement process on key performance indicators like patient treatment delay, a metric incorporating both the patient's ED boarding period and time to travel to their eventual IP destination.

\textbf{Keywords} --- Discrete-Event Simulation,~\emph{Psychiatric Inpatient Care}, Emergency Department, Patient Flow,
Boarding, Patient Transfer%
\end{abstract}%

\section{SIGNIFICANCE AND PRACTITIONER POINTS}
This paper addresses the unique use-case of a queueing network discrete event simulation applied to mental healthcare operations and patient flow. The methodology and model design of a large, heavily interconnected system given limited data availability will be of interest to researchers in the field of operations research and systems engineering. For practitioners overseeing psychiatric transfer, our simulation model provides a framework for identifying potential bottlenecks and inefficient processes preventing timely mental healthcare access. Additionally, the proposed interventions serve as a starting point for the development of potential systemic improvements designed to enhance access to psychiatric inpatient healthcare. 

\section{INTRODUCTION}
\label{Introduction}
% What is Emergency Department Boarding?
The issue of lengthy waiting times for patients is an established and well-researched topic in a wide variety of healthcare settings, such as primary care access \citep{Ansell2017}, surgical units \citep{Willcox2007}, and outpatient settings \citep{lewis2018}.
One of the most heavily researched concentrations is wait times in Emergency Departments (EDs) and their effect on ED operations. Researchers concur that the primary contributor to ED overcrowding, one of the most prevalent disruptive issues plaguing EDs, is boarding \citep{Blum2008}. ED boarding is the intentional holding of a patient in an ED after the decision to either admit a patient or transfer them to an alternative facility for inpatient treatment has been determined \citep{Falvo2007}. This phenomenon is a considerable factor in extended ED patient length of stay (LoS) and affects a patient's timely access to care, directly contradicting the timeliness component of healthcare quality established by the Institute of Medicine \citep{Reid2005}.

Patients who need acute psychiatric inpatient (IP) care commonly go to an ED as the first point of contact and are repeatedly identified as disparately affected by ED boarding and extended LoS, both in the rate patients are boarded, and how long of await they will experience \cite{Nicks2012, ONeil2016}. Adoption of new laws \citep{Flowers2018}, and closures of publicly funded psychiatric hospitals \citep{Misek2017, Torrey2012, Fuller2016} have amplified the boarding problem in recent years. This poses a dangerous prospect considering prolonged ED stays have been associated with exacerbating patients' symptoms and patients exiting the ED before proper assessment \citep{Bender2008}. 

While the consequences of boarding are considerable for all psychiatric patients, its prevalence varies by patient characteristics. Consider a psychiatric patient who is awaiting placement in a psychiatric inpatient bed, of which none are available at the current hospital. This patient often will be transferred to another hospital's inpatient unit for treatment. A transfer disposition introduces new steps and potential bottlenecks in the care process, both while a patient remains in an ED bed (e.g. social workers sending referral requests, coordinating travel with another facility) and once the patient has physically exited their original hospital (e.g. travel time). Studies have identified a higher likelihood and longer periods of ED boarding among transferred patients, which is amplified in certain patient groups such as patients younger than 18 years old and those 65 years and older, \citep{ONeil2016,Warren2016}. Additionally, delayed access to inpatient care is compounded by travel time to the new facility; valuable time where the patient is not receiving the targeted care they need. In light of this, it may be beneficial to consider overall treatment delay, which incorporates travel time in models rather than ED boarding time or ED LOS in isolation. This is the perspective we take in this study.

In the operations research and emergency medicine literature, researchers have examined patient characteristics associated with boarding and extended ED LoS \citep{Smith2016}, as well as novel changes that can be made within hospital EDs, to reduce treatment delays of patients. Improved staffing allocation \citep{Chen2020a} and bed management \citep{Lee2021} have been shown to help ameliorate care access and reduce waiting times. However, as discussed by Roh et al. \citet{Roh2019}, improving internal patient flows and operations in the ED can only do so much good. Instead, the primary goal of this study is to model patient flow out of the ED for psychiatric patients by simulating the internal admission and external transfer pathways for psychiatric inpatient patients including the process for determining a patient's final inpatient treatment location. Additionally, we propose and evaluate alternative designs to external transfer referral processes to reduce ED boarding and place patients in inpatient beds.

Researchers have used DES extensively in studies analyzing healthcare delivery \citep{Marshall2015,Marshall2015a} and ED operations \citep{Vanbrabant2019} although sparingly in the mental health context \citep{Noorain2019}. DES has been used to identify potential breakdowns in operations due to increasing trends in patient volume, and propose possible improvements to care facilities to prove patient care access such as dedicated ED services for those in crisis \cite{BaiaMedeiros2019}. Yet the current literature lacks a focus on the system-wide implications due to the interconnected nature of units across individual hospitals and stakeholder involvement. Proposed improvements in the current literature are restricted to individual care facilities and, as identified by La et al. \citet{La2016}, these improvements would require significant material investment to realize sizeable care access benefits. System-wide interventions could exhibit similar benefits and DES is an excellent tool to investigate this potential.

The remainder of this paper is formatted as follows; Section \ref{section:prob_statement} offers a brief description of the problem and its setting; Section \ref{section:DES} is a description of the simulation model, its input parameters, its validation and a description of proposed system-wide interventions aimed to improve access to psychiatric inpatient care; and Section \ref{section:Results} discusses the results of the baseline simulation and the interventions. Finally, we discuss the practical implications of the results in Section \ref{section:Discussion + Conclusion} section. This work was completed in collaboration with a team of emergency department and mental health physicians, social workers, and staff at a large academic medical center located in the midwest United States (which will be referred to as reference hospital for the remainder of this paper), and our model consists of care facilities in the state. This team was vital in providing both firsthand expertise on the subject matter and relevant data for modeling parametrization and validation. 

\section{INPATIENT PLACEMENT AND TRANSFER PROCESS}
To demonstrate the interdependencies between the various EDs and IP units, we consider two patient populations and their interactions with the aforementioned units. These groups are patients who are referred for IP care and admitted to the appropriate unit of the same hospital as the ED they arrived at (1) and patients who are referred and admitted to IP beds from any other source (2). For brevity, the word "psychiatric" will be omitted when referencing psychiatric patients, IP units/care, etc.% For simplification and modeling purposes patients in the 2nd population are assumed to request and occupy inpatient beds immediately.

The exact process for patients entering the care process at the ED differs depending on the hospital but most follow a similar general pathway. Patients receive both a physical (e.g., to test patient alertness, stability, etc.) and risk assessment during their ED stay. The risk assessment is especially crucial to help determine behavioral characteristics that may affect patient placement during the referral phase, such as a history of substance abuse, suicidal ideations, etc. After all requisite assessments, an ED physician will decide if IP care is necessary, and if so, the patient information is sent to a social worker for referral and placement. The ED physician may also decide that the patient does not require IP care, in which  case, the patient will be discharged and sent home.

The preferred placement choice is an IP bed at the same hospital as the ED the patient arrived. If an IP bed for the type of psychiatric patient under consideration is available when disposition is determined, the patient will be placed there. Licensing, hospital preferences, and operational rules---such as those related to a patient's age, severity, and history of violence---restrict the type of patients that may be placed in an IP unit's beds. If the original facility has no available beds or does not offer IP psychiatric care, the social worker must submit a transfer request to the nearest facility with available beds in accordance with the federal Emergency Medical Treatment and Labor Act \citep{emtala_1986}. 

While social workers are supposed to request transfer to the nearest hospital, there are often cases where that hospital rejects a majority of their received requests. Social workers will often notice which facilities are more likely to reject a patient's transfer requests, increasing the relevant patient's boarding period. Instead, patients may be better served if social workers were allowed to first request transfer to hospitals that are more likely to accept the patient or if they are allowed to send multiple transfer requests to different hospitals simultaneously rather than the current system of sending individual requests successively. We explore both these possibilities in the proposed interventions in Section \ref{interventions_section}.

\begin{figure}[t]
    \centering
    \includegraphics[width = .8\textwidth]{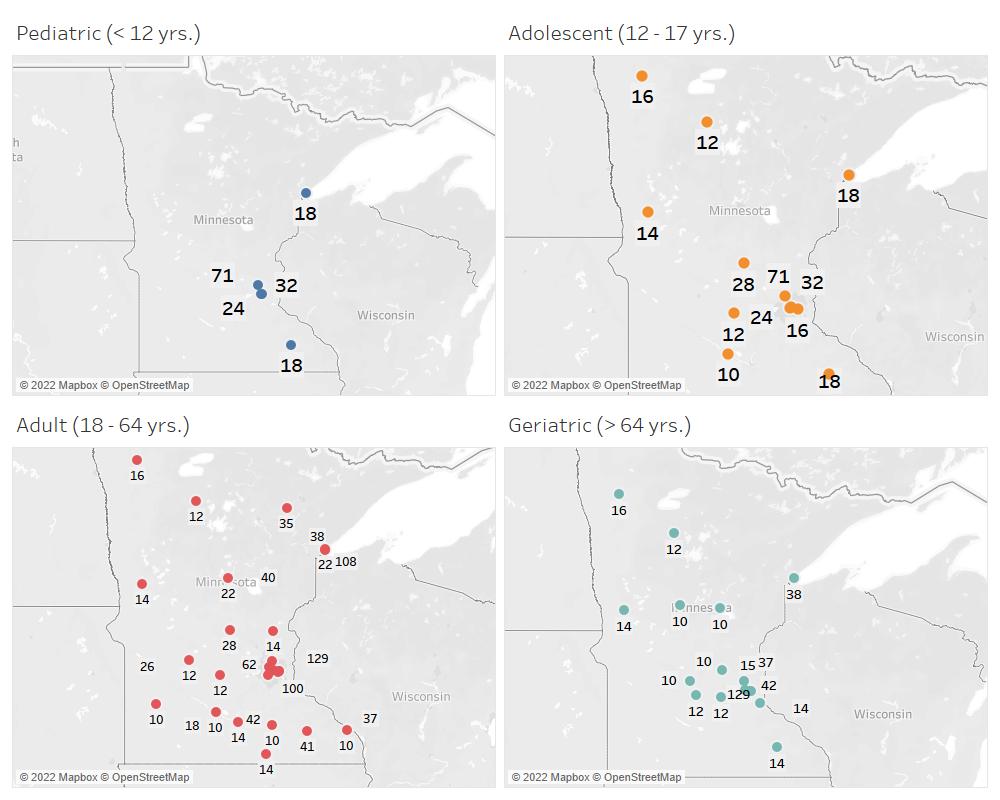}
    \caption{A map of the locations and capacities of inpatient psychiatric units across MN. Each value indicates the number of beds licensed for the corresponding age.}
    \label{fig:bed availability snapshot}
\end{figure}

An online bed tracking tool, made available by the Minnesota Hospital Association, gives social workers an indication of unoccupied IP beds in other hospitals \citep{bedtracker2022}. The IP units containing these beds are designated to treat patients in one or more of the four patient age groups: Children (ages 11 and younger), Adolescents (ages 12 - 17), Adults (ages 18 - 64), and Geriatric (ages 65 and older), and available beds are listed on the tracking website according to which age group they are licensed for. Hospitals are not mandated to update this tool with available beds and some facilities do not list available beds. Consequently, social workers may have to contact hospitals individually to inquire about the relevant unit's occupancy status. Both the number of beds (refer to Figure \ref{fig:bed availability snapshot}) and available treatment capacity at any given time differ significantly for different age groups; fewer beds across the state are licensed specifically for patients 17 years or younger or those 65 years old and older, and are rarely posted on the bed tracker as available for transfers. Meanwhile, more beds are licensed for patients between the ages of 18 and 64. Although access to IP psychiatric beds is also a challenge for adults, there are generally more beds available to adults at any given time compared to those available to child, adolescent, or geriatric patients.

Once the nearest hospital with available capacity is identified, the social worker will send a transfer referral along with relevant patient information including patient demographics, lab results, physician notes, etc. The requested facility's attending physician reviews the request and supporting material before making an admission decision. If the patient's transfer request is accepted, they are then transported to the facility to begin their treatment. Patients are usually driven to their new facility via ambulance but other modes of transportation (e.g. helicopter/driven in a private vehicle/driven by law enforcement) are also occasionally used. If the external facility rejects a request, the social worker locates the next nearest hospital with an available bed and repeats the process again until the patient is accepted somewhere. Because the decision to accept a transfer request is up to the transfer hospital's policies and the attending physician's discretion, more medically complex patients like those with so-called "placement barriers" identified during the risk assessment (e.g. history of violence or substance abuse) are likely to have multiple transfer requests rejected. Each referral and eventual rejection compounds the complex patient's ED boarding time. 

%Social workers experience facilitating patient transfer often gives them some intuition on which hospitals are likely reject a patient's transfer referral.  

\label{section:prob_statement}

\section{THE SIMULATION MODEL} \label{section:DES}
\begin{figure}[h]
    \centering
    \scalebox{.55}{\includegraphics{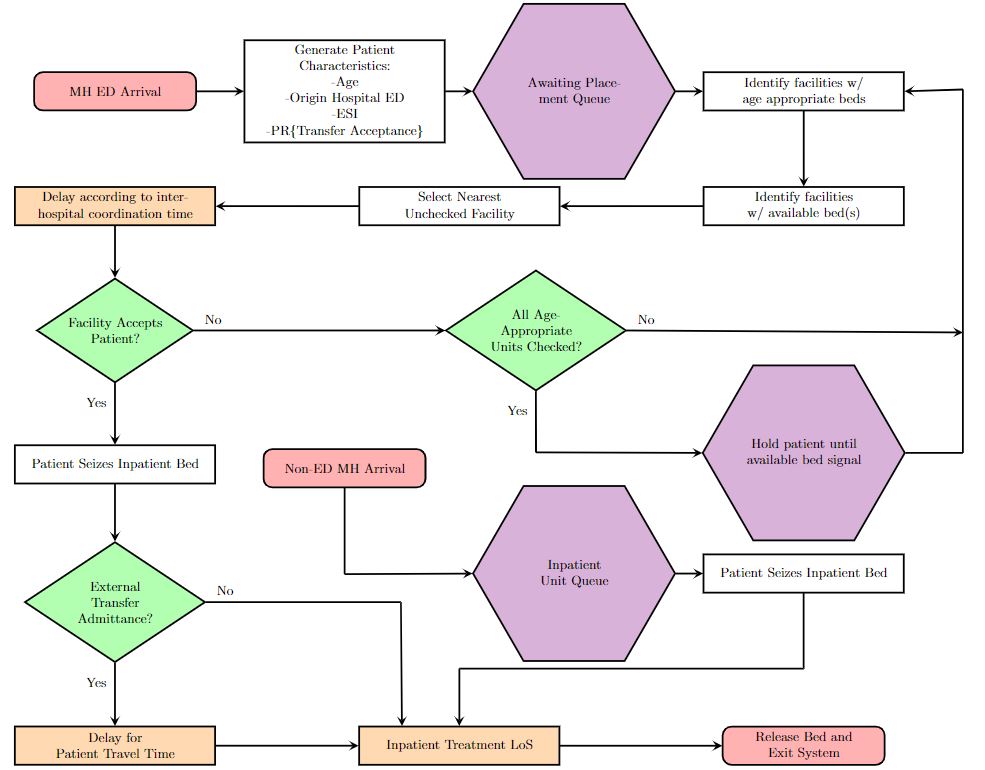}}
    \caption{This flowchart depicts the structure of our proposed DES model including the process for identifying which IP unit a patient is eventually transferred to.}
    \label{fig:flowchart}
\end{figure}

We developed a discrete event simulation to model patient flow from EDs to various IP units throughout the state. This simulation was coded using Simmer (version 4.4.3), a discrete event simulation library available for R \cite{Ucar2019}. The model runs for 365 simulated days with a 30-day warm-up period and for 20 replications. A flowchart of patient flow through the model can be found in Figure \ref{fig:flowchart} and is described in the following paragraphs.

 Our simulation model is a queueing network of individual EDs and facilities that contain up to 4 unique IP units with beds licensed for treatment for one or more age groups (the set of all IP units is denoted $\mathbb{F}$). Each ED contributes an arrival stream of patients requiring IP care and follows a Poisson arrival process with rate $\lambda_{ED}$. Each newly generated ED arrival is assigned an age group (Child, Adolescent, Adult, or Geriatric) indicating the type of IP psychiatric bed the patient is eligible to use. %, and an Emergency Severity Index (ESI) value indicating the patient's acuity and resource needs. 
Attributes are assigned according to their rates of occurrence in historical operational data. Additionally, patients are assigned a probability of transfer request rejection ($\alpha$), which allows us to capture the reality that it is more difficult for patients with certain characteristics to be admitted into an IP bed. After attribute assignment, the patient is assigned an IP bed and coordination time according to the process described in Section \ref{branch_func_description}. After an available bed is located and assigned to the patient, they are assigned an IP LoS corresponding to their assigned IP unit and are delayed according to the estimated drive time between their origin ED and their assigned IP unit ($\tau$). After the travel-time delay is exhausted, the simulated patient seizes the assigned IP bed resource, they are again delayed for the duration of their assigned IP LoS. Once their assigned LoS is exhausted, the patient releases their bed resource and exits the system.

 While this study is primarily concerned  with improving access for ED patients, it is necessary to include patients who access IP psychiatric beds without passing through the ED such as those transferred directly to a hospital's psychiatric IP unit from another IP unit and those referred directly from a clinic/physician's office or some other non-healthcare facility. In the simulation model, the non-ED patient arrival process is also Poisson with rate $\lambda_{IP}$. After being generated, these arrivals are assigned an IP LoS, seize an IP bed if one is available, and are delayed according to the assigned LoS. If no beds are available, the patient waits for a bed to become available in the IP unit's queue. Once a patient's LoS has exhausted, they release the bed and exit the system. 

\subsection{Input Parameters}

A key aspect of our model is the representation of IP units at multiple independent hospitals and hospital systems; thus, modeling parameters for each of these locations is a main requirement for the simulation. Table \ref{tab:Inputs_Table} briefly describes the input parameters we use and their respective sources. While ideal, accessing patient-level data from 150+ individual facilities, each with an ED and/or 1 or more IP psychiatric units, in order to calculate accurate arrival metrics is unrealistic. Instead, we approximate the facility's input parameters by combining anonymized patient-level data from our reference hospital's ED and IP units, and the results of the Minnesota Health Care Cost Information System (HCCIS), an annual survey of all medical providers in the state put forth by the Minnesota Department of Health \cite{hccisdata}. Survey results include annually aggregated operational statistics such as total patient admissions, total beds available, and total patient days for each hospital, and many of these measures are aggregated by sub-specialty, unit, etc.

Unique patient arrival rates for each considered ED are a necessary input parameter of the simulation. These were determined by first calculating the average number of daily arrivals to our reference hospital's ED who require IP care. This value varies by day and is significantly smaller during the weekend (Table \ref{tab:Rates_Table}). To reflect this, simulated ED patients follow a non-stationary Poisson process with different arrival rates for each day of the week. The data used to calculate this arrival rate pertains only to patients entering our reference hospital's ED and provides no information about the corresponding ED arrival rates for other hospitals. The aforementioned HCCIS dataset reports the total annual ED registrations for all MN hospitals, but this value alone cannot be used as a model input because it considers all ED patients regardless of diagnosis. To obtain estimates of other hospitals' ED-to-IP arrival rates, we assume the proportion of patients arriving at a hospital's ED that require IP psychiatric care is consistent across all EDs. Let $\mathbb{K}$ denote the set of all EDs in the simulation model ($k \in \mathbb{K}$) and $\mathbb{D}$ denote the set of days in a week ($d \in \mathbb{D}$). Under this assumption, we calculate the ratio of average daily ED arrivals at the reference hospital requiring IP treatment on day d ($n_{RF_{ED\xrightarrow{}IP}}^d$) to the daily ED registrations at the reference hospital, reflected in HCCIS ($n_{RF_{ED}}$) to obtain day $d$'s proportion of ED patients requiring IP psychiatric care ($\rho_{RF_{ED\xrightarrow{}IP}}^{d}$). This proportion is then multiplied by ED $k$'s reported daily ED arrivals from the HCCIS ($n_{k_{ED}}$) to get the estimated daily arrival rate ($\lambda_{k}^{d}$):

\begin{align}
    \rho_{RF_{ED \xrightarrow{} IP}}^{d} &= \frac{n_{RF_{ED \xrightarrow{} IP}}^{d}}{n_{RF_{ED}}} \: \forall \: d \: \in \mathbb{D}\\
    \lambda_{k}^{d} &=  \rho_{RF_{ED \xrightarrow{} IP}}^{d} \* n_{k_{ED}} \: \forall \: k \in \mathbb{K} \: , \: d \in \mathbb{D}
\end{align}

\begin{table}[h] 
\centering
\caption{The mean number of daily ED arrivals ($n_{RF_{ED}}$), arrivals needing IP care ($n_{RF_{ED \xrightarrow{} IP}}^{d}$) at the reference hospital, and arrival rate scaling parameter ($\rho_{RF_{ED \xrightarrow{} IP}})$. These values are used to calculate $\lambda_{k}^{d}\:\forall\: k\:\forall\:\mathbb{K},\:d\:\forall\:\mathbb{D}$} \label{tab:Rates_Table}
    \begin{tabular}{l c c c}
        \hline
        \textbf{Day} & \textbf{Average} & \textbf{95\% CI} & \textbf{$\rho_{RF_{ED \xrightarrow{} IP}}$}\\
        \hline
        \textbf{All Arrivals}\\
         \hspace{5mm} Total & 179.34 & -- & --\\
        \textbf{Requiring Psychiatric IP Care} & \\
         \hspace{5mm}Sunday & 2.31 & (2.11, 2.5) & 0.0128\\
         \hspace{5mm}Monday & 5.10 & (4.76, 5.45) & 0.0284\\
         \hspace{5mm}Tuesday & 5.04 & (4.71, 5.39)& 0.0281\\
         \hspace{5mm}Wednesday & 4.83 & (4.47, 5.19) & 0.0269\\
         \hspace{5mm}Thursday & 4.38 & (4.02, 4.75)) & 0.0244\\
         \hspace{5mm}Friday & 4.80 & (4.48, 5.13) & 0.0268\\
         \hspace{5mm}Saturday & 2.59 & (2.33, 2.83) & 0.0144\\
        \hline
    \end{tabular}
\end{table}

The IP unit arrival rate for patients who access IP care without an initial ED visit is approximated by scaling the arrival rate of this patient type into the reference hospital's IP unit, by the relative yearly volume of the IP unit of interest. Letting $\mathbb{F}$ denote the set of IP units in the model ($f \in \mathbb{F}$), we calculate unit $f$'s non-ED arrival rate ($\lambda_{f}$) by taking the product of the ratio of HCCIS reported daily arrivals for the reference hospital IP unit ($n_{RF_{IP}}$) to that of IP unit $f$ ($n_{f_{IP}}$) and the average daily non-ED IP arrivals seen in the reference hospital's operational data ($\lambda_{RF_{IP}}$) as shown in equation \ref{equation:non-ED_arrival_rate}.

\begin{equation} \label{equation:non-ED_arrival_rate}
    \lambda_{f} = \lambda_{RF_{IP}} * \frac{n_{f_{IP}}}{n_{RF_{IP}}}
\end{equation} 

Next, we must estimate how long patients spend in their IP beds/units. As identified in Roh et al. \citet{Roh2019}, the distribution of IP LoS is significantly skewed and does not fit theoretical distributions with reasonable accuracy, and to compensate, the authors simulated patients' LoS directly from the empirical distribution. We follow a similar direct sampling approach for IP LoS. However, in recognition that patient treatment time is not homogeneous among all IP units, we scale the sampled LoS. For any IP unit $f$, we assume the average LoS (in hours) to be IP unit $f$'s annual patient days ($g$) divided by its annual admissions ($a$). Each simulated patient's sampled LoS is scaled by the ratio of the reference hospital IP unit's average LoS (according to HCCIS) and the patient's destination IP unit's average LoS as seen in equation \ref{equation:mu_f}. 

\begin{equation}
    \label{equation:mu_f}
    \mu_{f} = \frac{g_{f} * 24}{a_{f}}
\end{equation}

The aforementioned reference hospital ED operational data also provides information for patients transferred to other facilities from EDs at the reference hospital and its affiliated facilities. This information includes each care facility contacted by the social worker while securing transfer, timestamps for every interaction with those facilities, and the patient's ultimate transfer destination. This data was used to determine estimates for parameters associated with facilitating patient transfer: each facility's likelihood of accepting a transfer request and the time needed to review a request and contact the social worker with a decision. Facility coordination times are assumed to be exponentially distributed with mean $\delta$. Timestamps for most facilities (specifically those who have available capacity when contacted by a social worker) occur in pairs: one for the initial contact ($t^{(1)}$) and one for the facilities transfer request response ($t^{(2)}$). To estimate the average coordination time $\delta$ for a facility $h$, let $\mathbb{P}$ be the set of all transferred patients ($p \in \mathbb{P}$), and $H_{p}$ be the subset of all facilities ($H_{p} \subseteq \mathbb{H}$) that a social worker contacts to a transfer patient $p$; we average the difference in contact times for each $h \in H_{p}$ over all transferred patients:

\begin{align}
    \delta_{h} &= \frac{\sum_{p \in \mathbb{P}} t^{(2)}_{p,h} - t^{(1)}_{p,h}}{\sum_{p \in \mathbb{P}} \mathbb{1}_{H_p}(h)} \: \forall \: h \in \: \mathbb{H} \\
    \mathbb{1}_{H_p}(h) &= 
    \begin{cases}
        1 & \text{if}\:h\in H_{p} \\
        0 & \text{o/w}
    \end{cases}
\end{align}

Simulated coordination times are then sampled from an exponential distribution with a rate parameter of $\delta_{h}^{-1}$.

A transfer facility's likelihood of accepting a patient ($\gamma_h$) is calculated in a similar manner. Each facility's second timestamp ($t^{(2)}$) is accompanied by a transfer decision (accept or reject), and so a facility's likelihood of acceptance is simply the ratio of accepted transfer requests to transfers requested:

\begin{align}
    \gamma_{h} &= \frac{\sum_{p \in \mathbb{P}}\mathbb{1}_{H_p}(p,h)}{\sum_{p \in \mathbb{P}} \mathbb{1}_{H_p}(h)} \: \forall \: h \in \mathbb{H} \\
    \mathbb{1}_{H_p}(p,h) &=
    \begin{cases}
        1 & \text{if patient}\:p\:\text{ is accepted by facility }\:h \in H_p\\
        0 & \text{o/w}
    \end{cases}
\end{align}

In both cases, we assume these parameters remain consistent regardless of a patient's origin location. In other words, a transfer request sent to facility $h$ will take, on average, $\delta_h$ hours if that patient originally came to the reference hospital's ED or any other emergency department. This assumption allows us to generalize these estimations for all simulated patients. 

\begin{table}[h]
    \centering
        \caption{A summary of various input parameters, their parameter/distribution or type of value, and their source.\\ (\textit{IP = Inpatient Unit, ED = ED, RF = Reference Hospital Data, HCCIS = Health Care Cost Information System})} \label{tab:Inputs_Table}
    \begin{tabular}{l c c c r}
        \hline
        \textbf{Input} & \textbf{Parameter} & \textbf{Unit Type} & \textbf{Distribution/Type} & \textbf{Source}\\ 
        \hline
        ED $k$ Daily Arrival Rate (on day $d$) & $\lambda_{k}^{d}$ & ED & Scaled Poisson & RF/HCCIS \\
        Unit $f$ Daily Arrival Rate & $\lambda_{f}$ & IP & Scaled Poisson & RF/HCCIS\\
        Unit $f$ \# of IP Beds & $b_{f}$ & IP & $\mathbb{N}$ & HCCIS\\
        Unit $f$ LoS Scaling Parameter & $\mu_{f}$ & IP & Scaled Empirical Dist. &  HCCIS \\
        Facility $h$ Transfer Request Review Time & $\delta_{h}$ & IP & Exponential & RF\\
        Facility $h$ Pr\{Accept Transfer Request\} & $\gamma_{h}$ & IP & $\mathbb{R} \in [0,1]$ & RF\\
        Patient $p$ Pr\{Rejection\} & $\alpha_{p}$ & IP & Triangular & RF\\
        Travel Time from ED $k$ to IP $f$  & $\tau_{f}^{k}$ & ED/IP & $\mathbb{R}$ (in hrs.) & Google Maps\\     
        \hline
    \end{tabular}
\end{table}

\begin{comment}
\begin{table}[H] \label{tab:characteristics}
\caption{Percentage of reference hospital ED arrivals eligible for each psychiatric beds licensed to each age category over a 2-year span (1/1/2019 - 12/31/2021)}
\centering
    \begin{tabular}{l c r} 
        \hline
        \textbf{Age Category} & \textbf{\% of Patients} & \textbf{N}\\
        \hline
        \hspace{3mm} Child ($< 12$) & 2.18 & 97\\
        \hspace{3mm} Adolescent ($12 - 18$) & 25.29 & 1125\\
        \hspace{3mm} Adults ($18 - 65$)  & 68.95 & 3067\\
        \hspace{3mm} Geriatric ($65+$)  & 3.57 & 159\\
        \hline
    \end{tabular}
\end{table}
\end{comment}

\subsection{Bed Assignment Policy} \label{branch_func_description}

% @Kayse: Does this need to be reworded, it makes sense to me but it might read unclearly
Each simulated patient who requires a transfer, either because their hospital of origin does not have an available psychiatric IP bed, or does not offer psychiatric IP treatment at all, follows the bed location process. In this process, a list of IP units that treat the patient's age and have available capacity at the current simulation timestamp is generated and ordered by proximity to the patient's origin ED. A coordination time is randomly sampled from the respective coordination time distributions for each unit in this list.

As previously stated, each IP unit has an associated acceptance threshold probability ($\gamma_{f}$) representing its likelihood to accept transfer referrals; similarly, each patient possesses a patient-specific transfer rejection probability ($\alpha$). We assume a patient will be accepted to the nearest facility $f$ for which $\alpha < \gamma_{f}$, denoted by $f^{*}$. If the ED a patient initially arrives at also has an accompanying psychiatric IP unit $f_o$ with an available bed, we set $\gamma_{f_{o}} = 1$ such that the patient will be treated at their current location and not be externally transferred. The patient's total coordination time is  the sum of the coordination times of facility $f^{*}$ and all nearer facilities for which transfer was requested.

In some instances, there will be no IP units for a patient to transfer into, primarily due to a lack of available capacity in all units that treat that patient's age, a patient's rejection probability is large enough ($\alpha \approx 1$) such that the patient is rejected by multiple units or a combination of the two. In either case, the simulated patient is held while awaiting a ``free bed signal". Individual IP units send these ``free bed" signals once a patient is discharged from IP treatment and releases the bed resource they previously held. There are four types of signals, one for each age category, and patients are only released to begin checking for available beds again when their corresponding ``free bed" signal is sent

\subsection{Model Verification and Validation} \label{validation + verification section}
This model was developed with continuous feedback from practitioners and social workers who work directly within the system to ensure model accuracy. This implementation of the patient routing process was agreed to be a suitable approximation of the true process.

% Validation procedures
To assess the validity of the DES model, select output metrics were compared to actual values from IP and ED operational data. Four metrics were considered for simulation validation:

\begin{enumerate}
    \item Mean patient coordination time (time from when the disposition is established until a bed is reserved for the patient)
    \item Median patient coordination time
    \item The average number of patients that are transferred into a reference hospital Psychiatric IP unit
    \item Average daily number of Mayo ED arrivals transferred to another facility 
\end{enumerate}

Only simulated patients who arrived at the reference hospital ED were considered for the first three validation metrics, while the final metric considers any simulated patient transferred to a reference hospital IP bed and did not originate from the reference hospital ED. Because the primary focus of this study is to assess and improve IP care access for patients who find access the most difficult, validation results are aggregated into two groups. The first group consists of patients either under 18 years old or 65 years old and older (henceforth referred to as vulnerable patients), while the second group consists of adult (18 - 64) patients. Adult validation results are segregated from others because there are significantly more beds licensed for the adult treatment and previous studies have indicated that IP access is less of a challenge for this patient cohort.   
Simulation inputs were tuned until true values fell within or around the 95\% confidence intervals of the simulated counterparts. Table \ref{table:verify1} shows the comparison of actual and simulated measures. We note that some simulated confidence intervals do not contain the target value; namely the mean treatment delay of internally placed vulnerable patients, the median treatment delays (excluding externally transferred vulnerable patients), and the daily transfer rates into/out of the reference hospital. However, in a majority of these cases, the difference between the upper or lower confidence interval bound and the target value is small (e.g. the largest discrepancy between true value and simulation confidence interval for average daily transfers represents an additional simulated patient about every 3 days). Due to both the size and complexity of the model, these differences can be regarded as negligible and we can consider the model valid (See Mans et al. \citet{Mans2010} for justification).

\begin{table}
\caption{A comparison of each of the metrics used for simulation validation against their true values.} \label{table:verify1}
    \centering
    \scalebox{.8}{
    \begin{tabular}{l c c}
        \hline
        \textbf{Validation Measure} & \textbf{Simulated 95\% CI} & \textbf{True Value}\\
        \hline
        \textbf{Mean Coordination Time (in hours):} & &\\
        Adult Patients & (0.54, 0.81) & 0.67\\
        Vulnerable Patients & (0.35, 0.44) & 0.37\\
        \hspace{5mm} - Internally Placed & (0.25, 0.29) & 0.34 \\
        \hspace{5mm} - Externally Transferred & (1.75, 3.03) & 2.57\\
        
        \textbf{Median Coordination Time (in hours):} & &\\
        Adult Patients & (0.12, 0.13) & 0.19\\
        Vulnerable Patients & (0.10, 0.11) & 0.19\\
        \hspace{5mm} - Internally Placed & (0.09, 0.10) & 0.18 \\
        \hspace{5mm} - Externally Transferred & (0.81, 1.23) & 1.02\\
        
        \textbf{Average Daily Transfers} & &\\
        Into the reference hospital IP & (1.59, 1.69) & 1.27\\
        \hspace{5mm} - Adult Patients & (0.43, 0.48) & 0.34\\
        \hspace{5mm} - Vulnerable Patients & (1.14, 1.22) & 0.93\\
        Out of the reference hospital ED & (0.73, 0.88) & 0.40\\
        \hspace{5mm} - Adult Patients & (0.64, 0.78) & 0.34\\
        \hspace{5mm} - Vulnerable Patients & (0.08, 0.12) & 0.05\\
        \hline
    \end{tabular}
    }
\end{table}

\subsection{Interventions} \label{interventions_section}

This study considers 3 interventions to improve patient access, each implemented during the referral process.

% Pick better subsection titles

% Post-hoc analysis tests might need to change due to test assumptions and how results should be interpreted
\subsubsection{Intervention 1: First Request Transfer to Facilities Most Likely to Accept} \label{alt_scem_1}
Social workers who have extensive experience facilitating patient transfer frequently have an intuition as to which hospital's are likely to accept transfer requests. Similarly, hospitals may be more likely to accept a transferred patient if they originate from another hospital in the same system and respond to transfer requests rapidly. Despite this, and in accordance with EMTALA, social workers still must submit requests to the nearest facility with available capacity. One could expect that allowing social worker discretion in the referral process would reduce patient treatment delays. Modeling a social worker's intuition of facility acceptance likelihood is difficult as each social worker's intuition differs. Thus, we allow full visibility of facility acceptance likelihood values within the simulation during the branching process. Basically, rather than ordering potential IPs by proximity to the patient, these facilities are ordered by their transfer acceptance likelihood, and then requests are submitted sequentially until an acceptance occurs.

\subsubsection{Intervention 2: Send Transfer Request to Different Facilities Concurrently} \label{alt_scem_2}
As previously discussed, social workers refer patients to an external hospital if their own hospital is unable to treat a patient, commonly due to a lack of capacity. Social workers refer patients one at a time, such that if a patient's referral is rejected, the time spent in the ED bed is effectively wasted. In light of this, the second intervention evaluates the effect of disseminating patient referrals to multiple hospitals concurrently. Concurrent referrals are sent in rounds with each social worker referring the patient to $m$ nearest hospitals (ranked in increasing order by drive time) within each round $r$. The increase in patient's coordination time during a referral round $r$ is the maximum of all referral review times incurred in round $r$ (provided all referrals during this round are rejected) or the referral time of the first accepted request. If all transfer requests in round $r$ are rejected, then another $m$ transfer requests are sent to the next $m$ nearest hospitals in round $r+1$. This process continues until a patient is either accepted or rejected by all potential hospitals.

\subsubsection{Intervention 3: Send Concurrent Transfer Requests to Facilities Most Likely to\\Accept the Patient} \label{alt_scem_3}
The final proposed scenario is a combination of the previous two alternatives. Similarly to Section \ref{alt_scem_1}, external facilities are sorted by the likelihood to accept a transfer request. However, rather than sequentially submitting requests, social workers submit multiple requests simultaneously as in Section \ref{alt_scem_2}.

\begin{comment}
        \textbf{Characteristics} &\\
        Any & 47.2\\
        2 or more & 13.5\\
        \hspace{3mm} Legal & 10.6\\
        \hspace{3mm} Substance Abuse & 9.5\\
        \hspace{3mm} Behavioral Discontrol  & 7.5\\
        \hspace{3mm} Special Requirements & 6.0\\
        \hspace{3mm} Aggression/Violence & 3.8\\
        \hspace{3mm} Developmentally Disabled & 2.8\\
        \hspace{3mm} History of Sexual Offense & 2.1\\
        \hspace{3mm} Medically Complex & 1.6\\
        \hspace{3mm} Family Complications & 0.6\\
        \hspace{3mm} Other & 2.6\\
\end{comment}

\section{RESULTS} \label{section:Results}
\subsection{Baseline Simulation Results} \label{baseline_results}
Due to the previously discussed disparities in psychiatric care access between Adult patients (18 - 64 years) old, patient-specific results are aggregated into vulnerable and Adult patient groups similar to how validation metrics are aggregated in Section \ref{validation + verification section}. Our primary measure of care access is the patient's treatment delay. This measure incorporates coordination time for locating and assigning a bed (for all patients) and, for patients transferred to a new facility, the time required to travel to their assigned bed. As expected, the Adult patient group ``outperformed" all other ages in treatment delay. On average adults had a shorter wait for inpatient beds (Mean = 1.56 hours, CI [1.54, 1.58]), while the vulnerable patient group experienced significantly longer waits (Mean = 2.83, CI: [2.78, 2.89]). 
%Interestingly, this disparity was reversed when looking at transferred patients where the mean treatment delay for the vulnerable patient group was 3.58 hours (CI: [3.52, 3.65]) while the adults were higher at 3.89 hours (CI: [3.83, 3.95]). However, while the adult transfers waited longer for an IP bed, they constitute a smaller portion of the total transfers. 
The baseline scenario's results support previous studies' \citep{ONeil2016} findings on the likelihood that vulnerable patients require external transfer. A larger percentage of simulated vulnerable patients were sent away from the original ED's hospital for treatment (85.1\%, [84.8, 85.4\%]) compared to their adult counterparts with only 58.8\% requiring external transfer (CI: [58.6, 59.2\%]).

Previous studies, such as \citet{ONeil2016}, have shown that transferred patients in the vulnerable group are sent further from their homes or origin EDs when compared to adults and our simulation's baseline results support this conclusion as well. Transferred adult patients consistently had a higher proportion of patients who were transferred to an IP unit within 10, 25, and 50 miles from their original ED (Table \ref{table:transfer distances}).

\begin{table}[h]
    \centering
    \caption{A larger proportion of adults were placed in an IP bed closer to their original arrival ED compared to the vulnerable patients}
    \begin{tabular}{l c c}
        \hline
         \textbf{Inpatient Unit Distance from ED} & \textbf{Percentage} & \textbf{CI}\\
         \hline
         \textbf{Within 10 miles} & &\\
         \hspace{5mm} Vulnerable Patients & 61.5 & (61.2, 61.8)\\
         \hspace{5mm} Adult Patients & 70.2 & (70.2, 70.6)\\
         \textbf{Within 25 miles} & &\\
         \hspace{5mm} Vulnerable Patients & 712 & (70.9, 71.6)\\
         \hspace{5mm} Adult Patients & 80.6 & (80.1, 80.6)\\      
         \textbf{Within 50 miles} & &\\
         \hspace{5mm} Vulnerable Patients & 82.1 & (81.8, 82.4)\\
         \hspace{5mm} Adult Patients & 92.1 & (92.0, 92.3)\\
         \hline
    \end{tabular}
    
    \label{table:transfer distances}
\end{table}

Additionally, we can look at the usage of inpatient beds throughout the state. Bed utilization in most units remained stable according to queuing theory ($<100\%$ bed occupancy), with an overall mean occupancy of 70.9\% across all IP units (CI: [70.6, 71.1\%]). This was consistent even when aggregating unit occupancy by their availability to certain age groups (Table \ref{table:age occupancy}). However, 7 of the 41 individual units showed an average IP bed occupancy over 90\% indicating they are often near full capacity and would be unable to handle a potential surge in IP demand. Conversely, 6 of the 41 units had an average occupancy level of less than 40\%. Interestingly, all of these units contain beds licensed for geriatric patient treatment (although 2 units allow for adult treatment as well).

\begin{table}[h]
    \centering
    \caption{The average occupancy level of inpatient units with beds licensed to each age group hovered around 70\% excluding geriatric patients}
    \begin{tabular}{l c c}
        \hline
        \textbf{IP Unit has bed for:} & \textbf{Average Occupancy (\%)} & \textbf{CI}\\
        \hline
        \textbf{All} & 70.9 & (70.6, 71.1)\\
        Children/Pediatric & 70.7 & (70.2, 71.2)\\
        Adolescent & 74.0 & (73.7, 74.3)\\
        Adult & 71.0 & (70.7, 71.3)\\
        Geriatric & 55.1 & (54.4, 55.8)\\
        \hline
    \end{tabular}
    \label{table:age occupancy}
\end{table}

\subsection{Sensitivity Analysis}
\begin{figure}[h]
    \centering
     \scalebox{.8}{\includegraphics{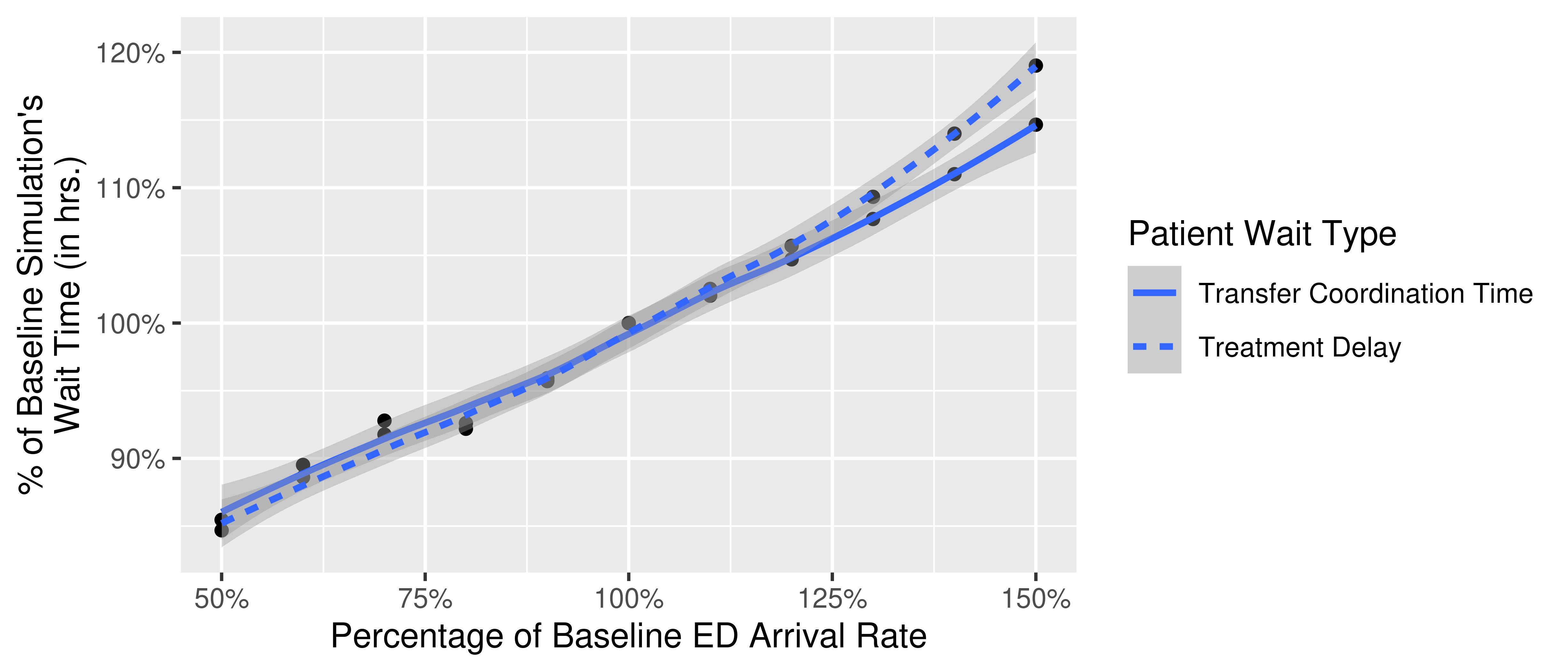}}
    \caption{Both the mean coordination time and treatment delay increased significantly as ED's arrival rates increased from 50\% to 150\% of their original rate.}
    \label{fig:sensitivity_analysis_1}
\end{figure}

To analyze how sensitive our chosen performance metrics are to the simulation inputs, two experiments were conducted. In the first experiment, we investigate how varying all ED's arrival rates affect both the patient's coordination time while a social worker finds an available bed and their overall treatment delay. We test this by running instances of the model with all ED arrival rates ($\lambda_k\:\forall\:k\:\in\:\mathbb{K}$) increasing from 50\% to 150\% of the baseline scenario's arrival rate. Simulated patients' wait times appear to increase linearly with the increasing ED arrival rate (Figure \ref{fig:sensitivity_analysis_1}). This should be expected because as the arrival rate increases, more patients are ``competing" to be placed in the same IP units/beds, but, because there is no change in IP unit characteristics (either in the total number of beds or LoS), bed availability remains about the same.

\begin{figure}
    \centering
    \scalebox{.8}{\includegraphics{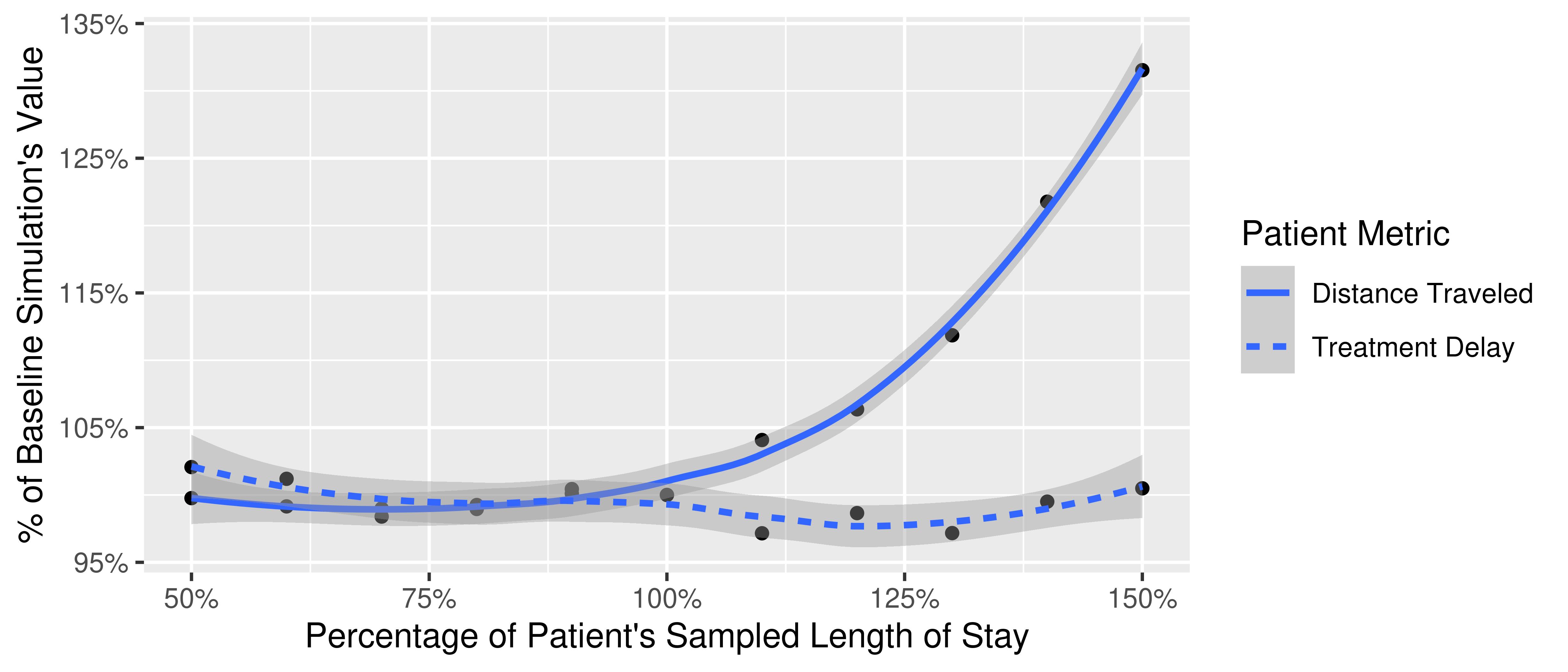}}
    \caption{The mean travel distance increases significantly when patient LoS was increased from 50\% to 150\% of their original value. The mean treatment delay remained unchanged throughout the range of length of stay values.}
    \label{fig:sensitivity_analysis_2}
\end{figure}

The second sensitivity analysis experiment investigates the effect of a patient's length of stay input on key performance metrics. Similar to the previous experiment, we run scenarios in which simulated patients' length of stay is a percentage of that of the baseline simulation, increasing from 50\% to 150\%. Interestingly, the varying LoS does not appear to have an influence on simulated patients' mean treatment delay and this measure remained relatively constant for all percentages amongst transferred vulnerable patients. This was not the case for the mean distance traveled by simulated patients, which increased significantly as the length of stay increased (Figure \ref{fig:sensitivity_analysis_2}). 

\subsection{Intervention Results}
Interventions 1 and 3 will be evaluated by the mean value of both coordination time and the treatment delay metric proposed in \ref{baseline_results}, among transferred patients in the vulnerable patient group. Both metrics are used for evaluation because the proposed interventions affect both the process of locating an available bed (and by extension the social worker coordination time) and the potential final treatment destination of the patient (which by extension, modifies their expected travel time). Intervention 2 will be evaluated solely by social worker coordination time for the same patient type. Interventions 2 and 3 are designed as single-factor experiments with the number of concurrent referrals within a round chosen as the treatment variable. Experiments for interventions 2 and 3 are conducted using the Kruskal-Wallis One-way Analysis of Variance and the Wilcoxon Rank-Sum test for pairwise comparisons between treatment levels. Comparisons between Intervention 1's results against that of the baseline scenario are conducted with the Mann-Whitney U-Test. Intervention 1 and each treatment level of Interventions 2 and 3 were run with 20 simulation replications with a 30-day warm-up period.

% Due to the heavy tails of the distribution of patient TTP under the baseline scenario, parametric experimental designs are not appropriate.

\subsubsection{Intervention 1 Results}
 The results of this intervention pertain only to transferred patients because this intervention modifies how transfer requests are sent and should not affect internally admitted patients. A t-test indicated a small but statistically significant reduction in the mean coordination time among vulnerable patients when compared to the baseline simulation results (Difference of means (hours): 0.151, CI: [0.062, 0.240], p-value = 0.0015). 

Conversely, the mean treatment delay appears to increase with the implementation of this intervention (Difference in Means: -0.084, CI: [-0.139 -0.0295], p-value = 0.0035) suggesting that patients traveled further to receive inpatient care under this policy. One should note, this model assumes patients are driven to an external facility and that this travel time is deterministic. In practice, both these assumptions may be violated and the increase in treatment delay may not materialize.

It is important that a proposed intervention does not negatively affect access to mental healthcare for those not included in the patient group of interest (any adult patient, and those internally placed). As expected, there was no significant shift in mean treatment delay for internally placed patients (Difference in Means (in hours): -0.005, CI: [-0.003  0.007], p-value = 0.411). Additionally, access to care for adult transfer patients did not change (Difference in Means (in hours): -.007, CI: [-0.040, 0.025], p-value = 0.649). 

\subsubsection{Intervention 2 Results} \label{section:int_2_results}
\begin{figure}[h]
    \centering
    \scalebox{.8}{\includegraphics{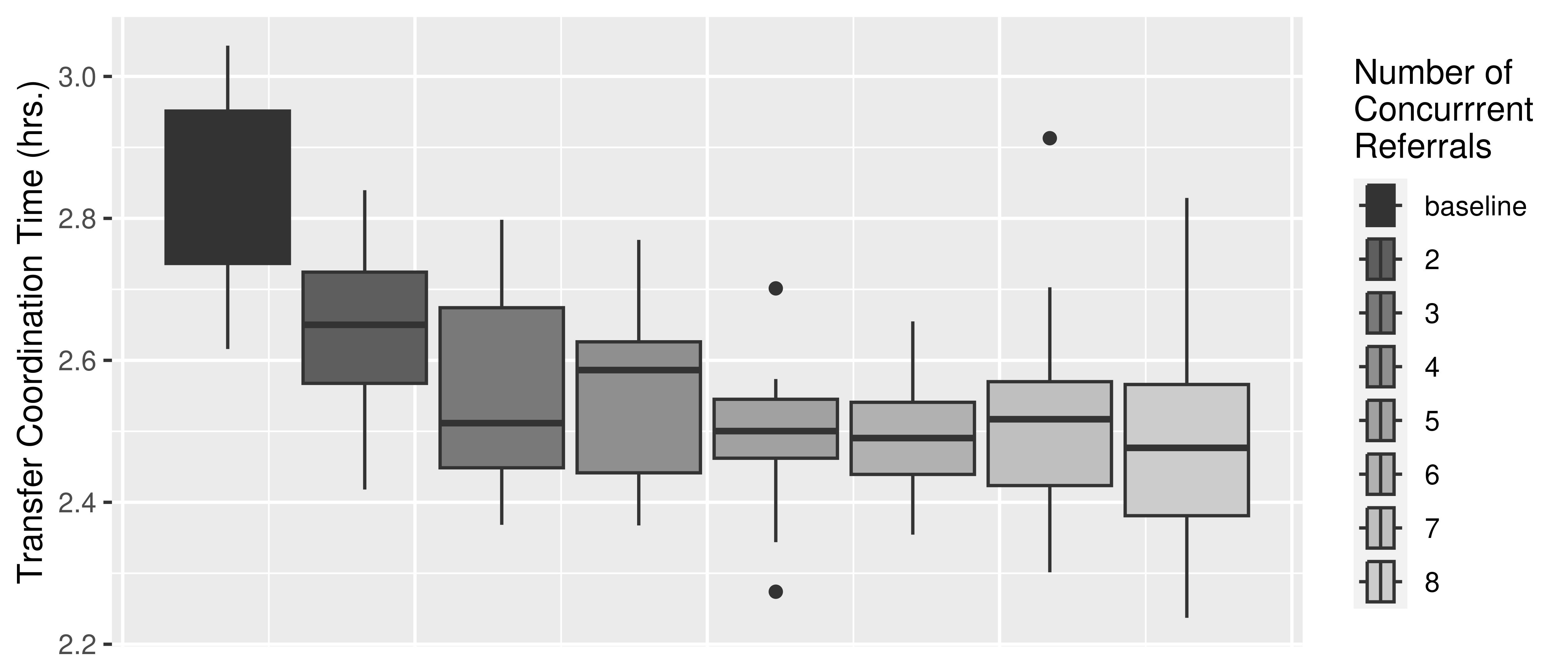}}
    \caption{Implementing any number of concurrent referrals decreases coordination time compared to the baseline simulation but the effect diminishes as more referrals are sent together.}
    \label{fig:int_2_boxplot}
\end{figure}

Intervention 2 showed a gradual decrease in the SW coordination time for locating and securing a bed for the patient. Kruskal-Wallis analysis of variance indicated a significant difference in mean coordination times between the different numbers of concurrent referrals sent. Pairwise comparisons of each concurrent referral scenario using Tukey's range test indicate that sending a minimum of two concurrent referrals will significantly decrease a patient waiting time. The  trend of decreasing coordination time when increasing the number of referrals is shown in Figure \ref{fig:int_2_boxplot}. 

\subsubsection{Intervention 3 Results}
\begin{figure}[h]
    \centering
    \scalebox{.8}{\includegraphics{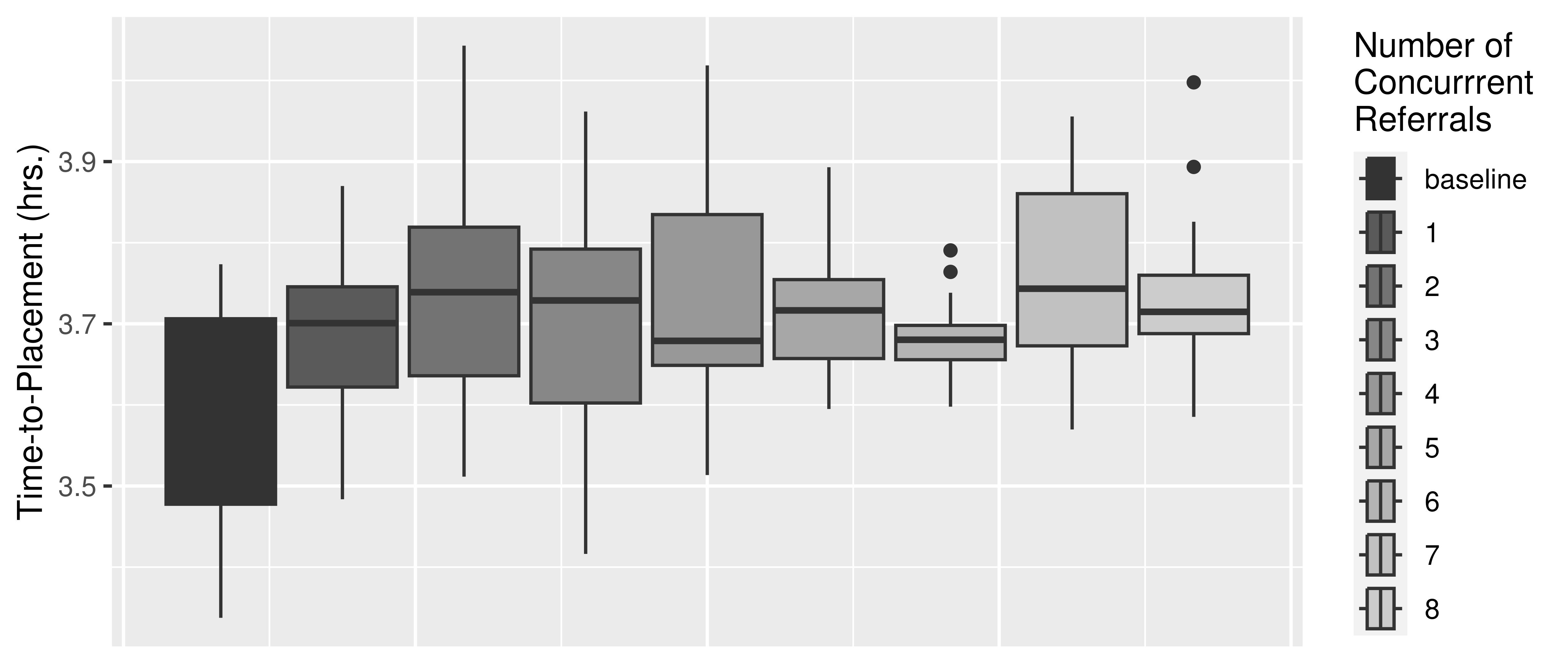}}
    \caption{Intervention 3 appeared to increase mean treatment delay, despite the decrease in coordination time compared to the baseline model.}
    \label{fig:int_3_boxplot}
\end{figure}

After implementing the third intervention in the simulation, results were similar to that of intervention 2 when examining the coordination time in isolation. However, the mean treatment delay did not differ significantly between each number of concurrent arrivals and exceeded that of the baseline model (Figure \ref{fig:int_3_boxplot}). This suggests that while the patient spends less time in the ED boarding in this intervention, this does not compensate for the extra travel a patient incurs by traveling to a further inpatient unit as was the case in intervention 1. 

\section{DISCUSSION AND CONCLUSION} \label{section:Discussion + Conclusion}
In this study, our goal is to model the IP bed assignment process for psychiatric patients presenting to the ED and reduce the patient's wait for IP treatment through proposed assignment policy alterations. Our baseline scenario results show the potential for extending the scope of healthcare discrete event simulation beyond the confines of a single hospital by modeling the flow of psychiatric patients into IP units from multiple EDs, both in the same hospital and other facilities. 

The results from Section \ref{section:Results} exhibit how structured DES experiments can be used to evaluate and improve mental healthcare policy. Overall, the proposed interventions show evidence of reducing the time social workers spend finding a hospital with available treatment capacity that is willing to accept an additional patient. Naturally, this decreased coordination time is beneficial for the patient as they wait for less time before knowing an IP bed is available for them. In some cases, this reduction is also beneficial for the social worker such as in Intervention 3, where fewer transfer referrals are sent before a patient is accepted for treatment. Treatment delay appeared to increase under Interventions 1 and 3, but these increases can be attributed to the travel time rather than coordination or ED boarding. These proposed interventions serve as a starting point to illustrate the benefit of a system-wide model like ours and further interventions in policy, resource availability, hospital operations, etc. should be tested for their efficacy in future research.

While the model and intervention's results are promising, some study limitations should be abated before insights and results from a model of this scope can be used to inform and make policy decisions. As described in Section \ref{section:DES}, many of the model's inputs were derived from either operational data of a single hospital and its interactions with surrounding hospitals and/or aggregated state-provided data. In the future, the fidelity of our proposed model could be improved by determining inputs from the operational data of all hospital units considered. Our model's logic has been generalized to apply to all the EDs and care providers considered, and in effect, ignores the different processes and nuances of individual care providers. Engagement with practitioners and social workers from these other facilities could help build a more robust and accurate model and should be explored in future studies. This would require stakeholder involvement from numerous and often competing hospitals, but the potential for both reduced ED boarding leading to a decreased strain on ED resources as well as improved access to IP care for many psychiatric patients is encouraging and the potential could prove to be an attractive incentive. 

\section{ACKNOWLEDGEMENTS}
We would like to thank our team of practitioners and collaborators at Mayo Clinic, St. Mary Campus: Nasibeh Zanjirani Farahani, David Nestler and Jennifer Condon for their assistance in this study.
\newpage

\bibliography{REFERENCES}

\begin{thebibliography}{10}

\bibitem{bedtracker2022}
Mental health service locator web site.
\newblock Minnesota Hospital Association, 2022.

\bibitem{Ansell2017}
Dominique Ansell, James~A.G. Crispo, Benjamin Simard, and Lise~M. Bjerre.
\newblock {Interventions to reduce wait times for primary care appointments: a
  systematic review}.
\newblock {\em BMC Health Services Research}, 17:1--9, 2017.

\bibitem{BaiaMedeiros2019}
Deyvison~T. {Baia Medeiros}, Shoshana Hahn-Goldberg, Dionne~M. Aleman, and Erin
  O'Connor.
\newblock {Planning Capacity for Mental Health and Addiction Services in the
  Emergency Department: A Discrete-Event Simulation Approach}.
\newblock {\em Journal of Healthcare Engineering}, 2019, 2019.

\bibitem{Bender2008}
David Bender, Nalini Pande, and Michael Ludwig.
\newblock {A Literature Review: Psychiatric Boarding}.
\newblock Technical report, The Lewis Group, 2008.

\bibitem{Blum2008}
Frederick~C Blum, West Virginia, Robert~I Broida, and Chief~Operating Officer.
\newblock {Emergency Department Crowding: High-Impact Solutions}.
\newblock Technical report, American College of Emergency Physicians, 2008.

\bibitem{Chen2020a}
Wenjie Chen, Hainan Guo, and Kwok~Leung Tsui.
\newblock {A new medical staff allocation via simulation optimisation for an
  emergency department in Hong Kong}.
\newblock {\em International Journal of Production Research}, 58:6004--6023,
  2020.

\bibitem{emtala_1986}
{Examination and Treatment for Emergency Medical Conditions and Women in
  Labor}.
\newblock 42 U.S.C. § 1395dd. 1986.

\bibitem{Falvo2007}
Thomas Falvo, Lance Grove, Ruth Stachura, David Vega, Rose Stike, Melissa
  Schlenker, and William Zirkin.
\newblock {The Opportunity Loss of Boarding Admitted Patients in the Emergency
  Department}.
\newblock {\em Academic Emergency Medicine}, 14:332--337, 2007.

\bibitem{Flowers2018}
Lee~M. Flowers, Kayse~T. Maass, Gabrielle Melin, Ronna~L. Campbell, Paul~J.
  Novotny, Jessica~J. Westphal, David~M. Nestler, and Kalyan~S. Pasupathy.
\newblock {Consequences of the 48-h rule: A lens into the psychiatric patient
  flow through an emergency department}.
\newblock {\em American Journal of Emergency Medicine}, 36:2029--2034, 2018.

\bibitem{Fuller2016}
Doris~A Fuller, Elizabeth Sinclair, Jeffrey Geller, Cameron Quanbeck, and John
  Snook.
\newblock {Going, Going, Gone. Trends and Consequences of Eliminating State
  Psychiatric Beds, 2016}.
\newblock Technical report, Treatment Advocacy Center, Arlington, VA, 2016.

\bibitem{hccisdata}
{Hospital Fiscal Year End Data}.
\newblock Accessed May. 10, 2022.

\bibitem{La2016}
Elizabeth~M. La, Kristen~Hassmiller Lich, Rebecca Wells, Alan~R. Ellis,
  Marvin~S. Swartz, Ruoqing Zhu, and Joseph~P. Morrissey.
\newblock {Increasing access to state psychiatric hospital Beds: Exploring
  supply-side solutions}.
\newblock {\em Psychiatric Services}, 67(5):523--528, 2016.

\bibitem{Lee2021}
Seung~Yup Lee, Ratna~Babu Chinnam, Evrim Dalkiran, Seth Krupp, and Michael
  Nauss.
\newblock {Proactive coordination of inpatient bed management to reduce
  emergency department patient boarding}.
\newblock {\em International Journal of Production Economics}, 231:107842,
  2021.

\bibitem{lewis2018}
Annie~K. Lewis, Katherine~E. Harding, David~A. Snowdon, and Nicholas~F. Taylor.
\newblock {Reducing wait time from referral to first visit for community
  outpatient services may contribute to better health outcomes: A systematic
  review}.
\newblock {\em BMC Health Services Research}, 18:1--14, 2018.

\bibitem{Mans2010}
Ronny~S. Mans, Nick~C. Russell, Wil {Van Der Aalst}, Piet~J.M. Bakker, and
  Arnold~J. Moleman.
\newblock {Simulation to Analyze the Impact of a Schedule-Aware Workflow
  Management System}.
\newblock {\em Simulation}, 86(8-9):519--541, 2010.

\bibitem{Marshall2015a}
Deborah~A. Marshall, Lina Burgos-Liz, Maarten~J. Ijzerman, William Crown,
  William~V. Padula, Peter~K. Wong, Kalyan~S. Pasupathy, Mitchell~K. Higashi,
  and Nathaniel~D. Osgood.
\newblock {Selecting a dynamic simulation modeling method for health care
  delivery research - Part 2: Report of the ISPOR dynamic simulation modeling
  emerging good practices task force}.
\newblock {\em Value in Health}, 18(2):147--160, 2015.

\bibitem{Marshall2015}
Deborah~A. Marshall, Lina Burgos-Liz, Maarten~J. Ijzerman, Nathaniel~D. Osgood,
  William~V. Padula, Mitchell~K. Higashi, Peter~K. Wong, Kalyan~S. Pasupathy,
  and William Crown.
\newblock {Applying dynamic simulation modeling methods in health care delivery
  research - The SIMULATE checklist: Report of the ISPOR simulation modeling
  emerging good practices task force}.
\newblock {\em Value in Health}, 18(1):5--16, 2015.

\bibitem{Misek2017}
Ryan~K. Misek, Ashley~D. Magda, Samantha Margaritis, Robert Long, and Erik
  Frost.
\newblock {Psychiatric Patient Length of Stay in the Emergency Department
  Following Closure of a Public Psychiatric Hospital}.
\newblock {\em Journal of Emergency Medicine}, 53, 2017.

\bibitem{Nicks2012}
B.~A. Nicks and D.~M. Manthey.
\newblock {The Impact of Psychiatric Patient Boarding in Emergency
  Departments}.
\newblock {\em Emergency Medicine International}, 2012:1--5, 2012.

\bibitem{ONeil2016}
Amy~M. O'Neil, Annie~T. Sadosty, Kalyan~S. Pasupathy, Christopher Russi,
  Christine~M. Lohse, and Ronna~L. Campbell.
\newblock {Hours and miles: Patient and health system implications of transfer
  for psychiatric bed capacity}.
\newblock {\em Western Journal of Emergency Medicine}, 17:783--790, 2016.

\bibitem{Reid2005}
Proctor~P. Reid, W.~Dale Compton, Jerome~H. Grossman, and Gary Fanjiang.
\newblock {\em {Building a better delivery system: A new engineering/health
  care partnership}}.
\newblock 2005.

\bibitem{Roh2019}
Thomas Roh, Valerie Quinones-Avila, Ronna~L. Campbell, Gabrielle Melin, and
  Kalyan~S. Pasupathy.
\newblock {Evaluation of interventions for psychiatric care: A simulation study
  of the effect on emergency departments}.
\newblock In {\em Proceedings - Winter Simulation Conference}, volume
  2018-Decem, pages 2507--2517. IEEE, 2019.

\bibitem{Smith2016}
Joseph~L. Smith, Alessandro~S. {De Nadai}, Eric~A. Storch, Barbara
  Langland-Orban, Etienne Pracht, and John Petrila.
\newblock {Correlates of length of stay and boarding in Florida emergency
  departments for patients with psychiatric diagnoses}.
\newblock {\em Psychiatric Services}, 67, 2016.

\bibitem{Torrey2012}
E.~Fuller Torrey, Doris~A Fuller, Jeffrey Geller, Carla Jacobs, and Kristina
  Ragosta.
\newblock {No Room at the Inn: Trends and Consequences of Closing Public
  Psychiatric Hospitals 2005-2010}.
\newblock Technical report, The Treatment Advocacy Center, Arlington, VA, 2012.

\bibitem{Ucar2019}
I{\~{n}}aki Ucar, Bart Smeets, and Arturo Azcorra.
\newblock {Simmer: Discrete-event simulation for R}.
\newblock {\em Journal of Statistical Software}, 90, 2019.

\bibitem{Vanbrabant2019}
Lien Vanbrabant, Kris Braekers, Katrien Ramaekers, and Inneke {Van
  Nieuwenhuyse}.
\newblock {Simulation of emergency department operations: A comprehensive
  review of KPIs and operational improvements}.
\newblock {\em Computers and Industrial Engineering}, 131(December
  2017):356--381, 2019.

\bibitem{Warren2016}
Mark~B. Warren, Ronna~L. Campbell, David~M. Nestler, Kalyan~S. Pasupathy,
  Christine~M. Lohse, Karen~A. Koch, Eduard Schlechtinger, Scott~T. Schmidt,
  and Gabrielle Melin.
\newblock {Prolonged length of stay in ED psychiatric patients: A multivariable
  predictive model}.
\newblock {\em American Journal of Emergency Medicine}, 34(2):133--139, 2016.

\bibitem{Willcox2007}
Sharon Willcox, Mary Seddon, Stephen Dunn, Rhiannon~Tudor Edwards, Jim Pearse,
  and Jack~V. Tu.
\newblock {Measuring and reducing waiting times: A cross-national comparison of
  strategies}.
\newblock {\em Health Affairs}, 26:1078--1087, 2007.

\end{thebibliography}

\section*{AUTHOR BIOGRAPHIES}

\noindent{\textbf{Nathan O. Adeyemi}} is a Ph.D. Candidate in Industrial Engineering at Northeastern University, Boston MA. He previously earned his bachelor's degree in Industrial Engineering at the University of Massachusetts Amherst in 2019. In his doctoral research, Nathan focuses on simulation and stochastic optimization and their use in addressing problems of access to mental healthcare. He is a member of the Operations Research and Social Justice Lab (ORSJ) headed by Dr. Kayse Maass. His email is adeyemi.n@northeastern.edu\\

\noindent{\textbf{Kayse Lee Maass}} is an Assistant Professor in the Mechanical and Industrial Engineering department at Northeastern University where she leads the Operations Research and Social Justice Lab. She earned her Ph.D. in Industrial and Operations Engineering at the University of Michigan and completed her postdoctoral studies in the Department of Health Sciences Research at the Mayo Clinic. Dr. Maass’s research focuses on the application of operations research methodology to social justice, access, and equity issues within human trafficking, mental health, housing, and food justice contexts. Her email is k.maass@northeastern.edu\\

\noindent {\textbf{Kalyan S. Pasupathy}} is an expert in systems science and informatics, focused on both advancing the science and translating knowledge to improve health, well-being, and care delivery, demonstrated through his academic and practice leadership roles. He has founded and directed research and practice transformation programs and has over 20 years of experience leading and pioneering efforts in transforming large healthcare institutions and social service organizations. He is a prolific writer and has published a number of scholarly articles and two editions of a book on health informatics. His email is kap@uic.edu\\

\noindent {\textbf{Amanda M. Graham}} is a Licensed Independent Clinical Social Worker (LICSW) who received her Master of Social Work degree at the University of Minnesota in 2011. She has been employed at Mayo Clinic since 2012. Amanda started her career in outpatient pediatrics and floating within inpatient practice and the Emergency Department. Amanda has worked within the Emergency Department setting for eight years where she practiced as a social worker and then transitioned to supervising the group for six years. She has a special interest in pediatric and adolescent mental health. In the Emergency Department, she has been working to find ways to support those families when they encounter system barriers to getting the help they are seeking.
\end{document}